%% file: Cosmological_Perturbations_out_of_the_Box_1-arxiv-v2.tex
\documentclass[11pt,a4paper]{article}
\usepackage{jcappub}

\usepackage[utf8]{inputenc}
\usepackage{tocloft}
\usepackage[dvipsnames]{xcolor}
\usepackage{cite}
\usepackage{amsmath,amscd,amsthm}
\usepackage{amsfonts}
\usepackage{amssymb}
\usepackage{graphicx}

\usepackage[cal=cm,scr=dutchcal]{mathalfa}

\usepackage{wrapfig}
\usepackage{subcaption}

\usepackage{array}
\usepackage{hyperref}
\usepackage{framed}

\theoremstyle{theorem}
\newtheorem{thm}{Theorem}[section]

\newtheorem{cor}[thm]{Corollary}

\theoremstyle{definition}

\theoremstyle{remark}

\input{globaldef}

\input{commenting}

\newcommand{\dmd}{\diamondsuit}
\newcommand{\ind}[3]{{\ensuremath{#1_{#2} \cdots #1_{#3}}}}
\newcommand{\dd}{d^\dagger}

\newcommand{\atbnd}{\right|_{\pr M}}
\renewcommand{\H}{\mathcal{H}}

\newcommand{\bn}{{\mathbf{n}}}
\newcommand{\bt}{{\mathbf{t}}}

\newcommand{\hgamma}{{\gamma^t}}
\newcommand{\hSigma}{{\Sigma^t}}

\newcommand{\bvphi}{\bar{\varphi}}

\newcommand{\ttt}{{t^t}}
\newcommand{\tnd}{{t^{nd}}}

\title{Cosmological perturbations out of the box \MakeUppercase{\romannumeral 1}}
\author{E. Şeyma Kutluk}

\affiliation{İstanbul Technical University,\\
Maslak, İstanbul, Turkey}

\emailAdd{kutlukes@itu.edu.tr}

\abstract{Using the tool of Hodge-Morrey decomposition of forms, we prove a new decomposition of symmetric rank-2 tensors on Ricci flat manifolds with boundary for arbitrary dimensions. Using this we reconstruct a new cosmological perturbation theory that allows for the scalar-vector-tensor type separation of the linearized Einstein equations with general boundary conditions. We discuss gauge transformations, gauge invariant quantities and as an example how the new decomposition works out in the single-field inflation scenario. For the scalar modes we get two copies of Mukhanov-Sasaki equation, one of them with a slight modification. Additionally we run a Weinberg-like argument for the existence adiabatic modes, and find some gauge-invariant solutions to the perturbations that exists whatever the constituents of the universe are.}
\begin{document}
\maketitle
\flushbottom

\section{Introduction}
It is a standard assumption of many cosmology work that the perturbations on the cosmological background vanish at infinity. A crucial place where this assumption is used is the scalar, vector, tensor (SVT) decomposition on spatial manifolds. Taking up a spherical topology, for example, a symmetric and traceless rank-2 tensor on a flat spatial slice decomposes as
\begin{equation}
t_{ij}= \lp \nabla_i \nabla_j - \frac{g_{ij}}{m} \nabla^2 \rp \lambda + \nabla_{(i} \psi_{j)} + t_{ij}^{TT}
\end{equation}
where $m$ is the dimension of the manifold and
\begin{equation}
\nabla^i \psi_i=0 \cg \nabla^i t_{ij}^{TT}=0 \gag g^{ij} t_{ij}^{TT}=0 \ .
\end{equation}
This assumption, thus the decomposition, no longer holds if one is interested in non-vanishing behavior. An example of this can be asymptotic symmetry discussions where fields do not vanish trivially, or any scenario one would like to consider a patching of the spacetime. \\

A decomposition of the completely antisymmetric tensors (forms) with more general boundary conditions is available in the form of Hodge-Morrey theorem \citep{Gunter}. This provides an analogue of the Hodge decomposition, on manifolds with boundary, where tensors can have off-the-boundary components. Note that with this description one can study any type of boundary condition. However to be able to use this for equations that involve symmetric tensors such as Einstein's equations one needs to have an analogue theorem for symmetric tensors. 

In the following, using the Hodge-Morrey theorem, we show the existence and uniqueness of a new decomposition for symmetric rank-2 tensors on manifolds with boundary via an argument similar to that of \citep{Straumann}. We start with an exposition of the concept of manifolds with boundary, some useful definitions on it and Hodge-Morrey theorem, mainly following \citep{Gunter}, in section \ref{sec:Man-bnd} \footnote{See also \citep{grmanton} for a summary of some concepts.}. In section \ref{sec:Dec-Thm} we prove the decomposition theorem \ref{thm:SVT} for symmetric rank-2 tensors on Ricci flat Riemannian manifolds with boundary. This theorem however turns out not to be good enough for the purposes of discussing gravitational perturbations: perturbed Einstein equations mixes some scalars with tensors of this decomposition unlike the standard treatment (such as \citep{MFB}). Because of this, in section \ref{sec:sec-dec}, we develop a second decomposition that facilitates the desired separation.

Having developed the full decomposition, in section \ref{sec:cos-per} we use it for cosmological perturbation theory, and decompose Einstein equations linearized around a cosmological background by discussing how various terms that appear in the equations are decomposed. We discuss gauge transformation of each component of the decomposition and define gauge invariant quantities. Finally in section \ref{sec:app-to-som}, we consider two scenarios for the cosmological perturbations: first we discuss the single inflation case and derive two-copies of Mukhanov-Sasaki equation, after this we discuss the existence of some gauge-invariant quantities independently of the background using an argument similar to that of Weinberg's \citep{weinbergadiabatic}.

\section{Manifolds with boundary and Hodge-Morrey decomposition of forms}\label{sec:Man-bnd}
A manifold with boundary can be defined in a similar way to a manifold, where the requirement of homeomorphism of open subsets to the open subsets of $\R^m$ is extended to homeomorphism to (relatively) open subsets of half-space $\R^+ \equiv \lcb x \in R^m | x^m \ge 0 \rcb$. On these manifolds derivatives of functions (and thus of tensors) on the boundary are defined as derivatives of smooth extensions of these functions evaluated at the boundary (\citep{Gunter},\citep{LeeISM}). In accordance with this definition, tangent vectors are defined such that on the boundary, vectors off $M$ are included e.g. $\left. TM \right|_{\pr M} \neq j_* \lp T\pr M \rp$, where $j: \pr M \ra M$.\\

From here on we will let M be a $\pr$-manifold: a smooth, orientable and metric complete manifold with boundary \footnote{See \citep{Gunter} for a complete definition.}. We will assume $M$ is equipped with a metric $g$ and the induced metric $j^* g$ on $\pr M$. On the boundary one can write down a unit outward pointing vector $N$ orthogonal to boundary i.e. a vector field that satisfies
\begin{equation}
g(N,N)=1 \gag g\lp j_* w, N \rp =0 \quad \mbox{everywhere on $\pr M$,} \ \forall w \in \Gamma(T\pr M) \ .
\end{equation}
Collar theorem (see e.g.\citep{Gunter}) then tells us that this $N$ can be extended into $M$ in a neighborhood of the boundary. From now on we will use the letter $N$ for denoting the normalized extension of the normal vector.
Because of the collar theorem, we can write any vector field $Y$ in the neighborhood of the boundary such that 
\begin{equation}
Y= g (Y,N) N + Y^\| \gwg g(Y^\|, N ) = 0 \ .
\end{equation}
Now with this decomposition, we define two objects that describes the orthogonal decomposition of any form on the boundary. For any form $w$, $\bt w$ is defined to be a map $\Gamma(\left. TM \right|_{\pr M}) \times \cdots \times \Gamma(\left. TM \right|_{\pr M}) \ra C^\infty (M)$ such that
\begin{equation}
\bt w \lp X_1, \cdots X_p \rp = w \lp X_1^\|, \cdots X_p^\| \rp \quad \forall X_i \in \left. TM \right|_{\pr M} \ ,
\end{equation}
whereas
\begin{equation}
\bn w := \left. w \atbnd - \bt w \ .
\end{equation}
Some identities in relation to these definitions follows:
\begin{align}
& *\bn w=\bt *w \gag *\bt w=\bn *w \ , \mbox{where $*$ is defined on $M$} \ , \label{eq:star-t identity}\\
&  j^* \bt dw= d j^* \bt w  \gag j^* * \bn \dd w = (-1)^{(k+1)(m+1)} d j^* * \bn w \ , \label{eq:star-d identity}
\end{align}
where $m:= \dim(M)= \dim(Int(M))$.
\noindent
We now move onto the decomposition theorem for k-forms. On compact manifolds (without boundary), the famous Hodge theorem establishes isomorphism between the set of harmonic forms and de-Rham cohomology groups, both of which is finite dimensional. On the contrary set of harmonic forms is infinite dimensional on manifolds with boundary. In place of them one defines harmonic, Dirichlet and Neumann fields:
\begin{align}
& 1) \ \Omega_D^k(M) = \lcb w \in \Omega^k(M) | \ \bt w=0 \rcb \cg \Omega_N^k (M)= \lcb w \in \Omega^k(M) | \ \bn w=0 \rcb \nonumber \\
& 2) \ \mbox{Harmonic fields:} \ \H^k(M)= \lcb w \in \Omega^k(M) | \ dw=0 \ \& \ \dd w=0 \rcb , \nonumber \\
& 3) \ \mbox{Dirichlet fields:} \ \H^k_D(M) = \H^k(M) \cap \Omega^k_D(M) \cg
\mbox{Neumann fields:} \ \H^k_N(M) = \H^k(M) \cap \Omega^k_N(M)  . \nonumber
\end{align}
On a compact $\pr$-manifold, $\H^k_{D/N}$ are finite dimensional and the correspondence to the de-Rham cohomology groups are as follows:
\begin{equation}
\mathcal{H}^k_D(M) \sim H^k(M,\dd), \mathcal{H}^k_N(M) \sim H^k(M,d) \ .
\end{equation}
Here one might wonder how to calculate the de-Rham cohomology of a manifold with boundary. If $M$ is a manifold with boundary then $M$ and $Int(M)$ are homotopy equivalent\footnote{ Theorem. 9.25 in \citep{LeeISM}.} and thus their de Rham cohomologies are the same at all orders. \footnote{Theorem 17.11 in \citep{LeeISM}.} 
Thus for example, de-Rham cohomologies of a ball with boundary is equivalent to that of an open ball of same dimension.
Now we state the decomposition theorem:
\begin{thm}[Hodge-Morrey Decomposition]\citep{Gunter}\label{thm:HM}
Any square integrable k-form on a compact $\pr$-manifold $M$ can be uniquely written as
\begin{equation}
w=d\alpha + \dd \beta + \kappa
\end{equation}
where $\kappa \in \H^k(M), \bt \alpha=0 , \bn \beta=0$ . 
\end{thm}
To illustrate the difference between vanishing boundary conditions and our investigation where by considering a manifold with boundary we allow for arbitrary boundary condition, let us consider the decomposition of a vector-equivalently a one form. In flat 3-dimensional space with trivial cohomology the Hodge-Morrey decomposition of a vector $V$ is
\begin{equation}\label{eq:vec_dec}
V^i = \nabla^i \alpha + \lp \vec{\nabla} \times \eta \rp^i + \nabla^i \sigma \ .
\end{equation}
where 
\begin{equation}
\bt \alpha=0 \cg *\bt \eta=0 \cg \nabla^2 \sigma=0 \ .
\end{equation}
Here we used the facts that in 3d $\beta=*\eta$ for some one-form $\eta$ and $d \kappa =0$ implies $\kappa=d \sigma$ for trivial cohomology. One can similarly write $\kappa= \dd \rho$, and thus absorb the harmonic part to exact or co-exact part if there were no boundary conditions. Thus we see that the ``standard" decomposition into exact and co-exact parts that is usually used for fields with vanishing boundary condition is not unique on a manifold with boundary if one does not include the boundary conditions.

\section{A decomposition theorem for symmetric rank-2 tensors}\label{sec:Dec-Thm}
The Hodge-Morrey decomposition works only for completely anti-symmetric tensors, but for the applications in gravity and cosmology one needs a similar decomposition for symmetric rank-2 tensors. Now we will perform this. We will consider symmetric rank-2 tensors that are traceless, on manifolds with boundary of arbitrary dimension that has trivial cohomology. (Trace can always be separeted uniquely.) \footnote{With trivial cohomology we always mean a manifold $M$ with boundary whose interior is contractible to a point. With Poincare lemma and homotopy equivalence of $M$ and $Int(M)$ this implies $H^k(M,d)=0$ for $k=1, \cdots,n$. } As in \citep{Straumann}, we start with the decomposition of $\nabla^i t_{ij}$. As a one form it has a unique decomposition
\begin{equation}\label{eq:div-t}
\nabla^i t_{ij} = \nabla_j \alpha + \dd \beta _j + \kappa_j
\end{equation}
where $\bt \alpha=0$, $\bn \beta=0$ and $\kappa$ is a harmonic field. Now the argument is, given $\dd \beta$ one can find a unique exact form $\tnd$ such that
\begin{equation}
\dd \beta = \dd d \tnd \gag \bn d\tnd=0 \cg \bn \tnd=0 \cg \dd \tnd=0 \ .
\end{equation}
This follows from corollary \ref{cor:ddw} and corollary \ref{cor:dw} stated in the Appendix; corollary \ref{cor:ddw} states that there exists a unique solution to the boundary problem
\begin{equation}
\dd \omega= \chi \cg d\omega=0 \cg \bn \omega=0 
\end{equation}
where $\chi$ is a given one form that satisfy $\dd \chi=0$, which we take to be equal to $\dd \beta$. Now we furthermore argue, given $\omega$ there exists a unique $\tnd $ such that 
\begin{equation}
d\tnd= \omega \cg \dd \tnd = 0 \cg \bn \tnd=0 \ .
\end{equation}
One can easily see that this follows from corollary \ref{cor:dw}. 
Now we proceed with the proof of the decomposition. First we define the traceless Hermitian 
\begin{equation}
\dmd_{ij}=\nabla_{(i} \nabla_{j)} - \frac{g_{ij}}{m} \nabla^2 ,
\end{equation}
where $m$ is the dimension of our $\pr$-manifold, for further convenience. A crucial step is that for a given one-form $V$
\begin{equation}\label{eq:phi}
(\dd d V)_j = \nabla^i \nabla_{[i} V_{j]} = \nabla^i \lp \nabla_{(i} V_{j)} - g_{ij} \nabla \cdot V \rp - R_j^i V_i .
\end{equation}
Thus we see that for a Ricci flat geometry one can write the coexact part as
\begin{equation}\label{eq:div-t}
(\dd \beta)_j =  \nabla^i \lp \nabla_{(i} t^{nd}_{j)} \rp 
\end{equation}
where the aforementioned conditions fix $t^{nd}$ uniquely. For the exact and harmonic parts, we make the following observation: with our assumptions a harmonic field $\kappa$ can be written as $\kappa=d\sigma$ for some $\sigma$. Then we note for a given scalar $\alpha + \sigma$ there exists a unique $\ttt$ such that
\begin{equation}
\nabla^2 \ttt = \frac{m}{m-1} (\alpha + \sigma) \gag \bt \ttt = 0 .
\end{equation}
This follows from Theorem 3.4.10 in \citep{Gunter} \footnote{Note that a 0-form $\alpha$ by definition satisfies $\bn \alpha=0$}. Given $\nabla^i t_{ij}$, $d(\alpha+\sigma)$ is fixed by the Hodge-Morrey theorem, thus $\alpha + \sigma$ is fixed up to a constant. However for a Ricci flat case
\begin{equation}
\nabla^i \dmd_{ij} \ttt=\nabla^i \lp \nabla_{(i} \nabla_{j)} \ttt - \frac{g_{ij}}{m} \nabla^2 \ttt \rp = \frac{(m-1)}{m} \nabla_j \nabla^2 \ttt ,
\end{equation}
thus the part coming from this constant does not contribute to the expression $\nabla^i \dmd_{ij} \ttt$. Because of this we see that if we define $t^{TT}$ such that
\begin{equation}\label{eq:existence}
t_{ij}= \dmd_{ij} \ttt + \nabla_{(i} t^{nd}_{j)}  + t^{TT}_{ij} \ ,
\end{equation}
then 
\begin{equation}
\nabla^i t^{TT}_{ij}=0 \gag g^{ij}t_{ij}^{TT}=0 
\end{equation}
and each term in the decomposition is unique. Uniqueness can alternatively be seen by proving if left hand side of \eqref{eq:existence} is zero, then each term in the right hand side is also zero which follows by using the arguments above. Let us now express the result as a complete statement:
\begin{framed}
\begin{thm}\label{thm:SVT}
Any symmetric traceless rank-2 tensor $t_{ij}$ on an arbitrary dimensional $\pr$-manifold with Ricci flat metric can be uniquely decomposed as
\begin{equation}
t_{ij}= \dmd_{ij} \ttt + \nabla_{(i} t^{nd}_{j)}  + t^{TT}_{ij} \ ,
\end{equation}
where
\begin{equation}
\bt \ttt=0 \cg \nabla^i \tnd_i=0 \cg \bn \tnd=0 \cg \bn d\tnd=0 \cg \nabla^i t_{ij}^{TT}=0 \cg g^{ij}t_{ij}^{TT}=0 .
\end{equation}
\end{thm}
\end{framed}

\section{A second decomposition}\label{sec:sec-dec}
In the next section we will see for applications to cosmological perturbation theory (though it will also be the case for any other gravitational perturbation theory), one will have mixing of the terms of the form
\begin{equation}
\nabla_i \nabla_j \sigma + t_{ij}^{TT} \ .
\end{equation}
where $\nabla^2 \sigma=0$. Note that $\nabla_i \nabla_j \sigma$ is transverse-traceless and thus cannot be separated from any other $t^{TT}_{ij}$ type of tensor with the decomposition above. In the next section however we will see such decomposition to be desirable. To perform such a decomposition we should ask ourselves the question: Given a transverse traceless symmetric rank-2 tensor $V_{ij}$, what other condition should it satisfy so that $V_{ij}=\nabla_i\nabla_j \sigma$ where $\pr^2 \sigma=0$. We study this question in flat space with the tools of \citep{Chodos}, where solutions to the Laplace equation ($\pr^2\sigma=0$) are studied using Taylor expansion.\footnote{See also \citep{Hui-Love}. We will only need the scalar part of the tensor spherical harmonic discussion in these references.} Given a rank-2 symmetric tensor inside a bounded region in $\R^m$, one can Taylor expand it so that
\begin{equation}
V_{ij}=\mathcal{V}_{ij|i_1 \cdots i_\ell} x^{i_1} \cdots x^{i_\ell} \gwg i_1,..i_\ell=1, \cdots m \ ,
\end{equation}
$\mathcal{V}_{ij|i_1 \cdots i_\ell}$ is separately symmetric in its first two and last $\ell$ indices and $(x^1,\cdots,x^m)$ are flat coordinates. Then $V_{ij}$ is transverse traceless if and only if
\begin{equation}
\mathcal{V}_{ii|i_1 \cdots i_\ell}=0 \gag \mathcal{V}_{ij|i_1 \cdots i_{\ell-1}j}=0 \quad \forall \ell \ .
\end{equation}
Now let us assume $V_{ij}=\pr_i\pr_j\sigma$, where $\pr^2 \sigma=0$. Note that $\sigma$ can be written as
\begin{equation}
\sigma=c_{i_1 \cdots i_\ell} x^{i_1} \cdots x^{i_\ell} \gwg c_{iii_1 \cdots i_\ell}=0 
\end{equation} 
and $c_{i_1 \cdots i_\ell}$ is completely symmetric in all of its indices. Then $V_{ij}=\pr_i\pr_j\sigma$ if and only if
\begin{equation}
\mathcal{V}_{ij|i_1 \cdots i_\ell} = (\ell+ 2) (\ell+1) c_{iji_1 \cdots i_\ell} \ .
\end{equation}
This means $\mathcal{V}_{ij|i_1 \cdots i_\ell}$ must be symmetric in all of its indices e.g.
\begin{equation}
\mathcal{V}_{ii_1|ji_2 \cdots i_\ell}=\mathcal{V}_{ii_2|i_1 j \cdots i_\ell}= \cdots = \mathcal{V}_{ii_m|i_1 \cdots j \cdots i_\ell} ,
\end{equation}
which in turn means
\begin{equation}
\left. \pr_{i_1} \cdots \pr_{i_\ell} V_{ij} \right|_{\vec{x}=0}= 0 \quad \forall \ell \ .
\end{equation}
By Taylor expanding $\pr_k V_{ij}$ one can show that this is equivalent to saying
\begin{equation}
\pr_i V_{jk}(x)=\pr_j V_{ik}(x) \quad \forall i,j,k \gag \forall x \in \R^m \ .
\end{equation}
This statement then is enough for the following decomposition theorem:
\begin{thm}\label{thm:sec_dec}
In the flat $\R^m$ space any rank-2 symmetric transverse-traceless tensor $V$ can be decomposed as
\begin{equation}
V_{ij}=X_{ij}+\pr_i \pr_j \sigma
\end{equation}
where $\pr^2\sigma=0$ and $X$ is a rank-2 symmetric transverse-traceless tensor that satisfies
\begin{equation}\label{eq:TTA}
\pr_{(i}X_{jk)}(x)=0 \quad \forall i,j,k \gag \forall x \in \R^m \ .
\end{equation}
We will use the notation that a rank-2 symmetric transverse-traceless that satisfies \eqref{eq:TTA} is denoted by the superscript $TTA$, e.g. $X_{ij}^{TTA}$.
\end{thm}

\section{Cosmological perturbation theory on spatial slices with boundary } \label{sec:cos-per}

With the new decompositions theorem \ref{thm:SVT} and theorem \ref{thm:sec_dec} at our disposal, we would like to see how Einstein's equations linearized around the FRLW background, which describes cosmological perturbations, decompose. As in the usual case with vanishing boundary conditions, first we will perform a space-time splitting. In this description all the tensors will be reconsidered to be a set of corresponding tensors on spatial slices with time parameters. For example a vector field $V^\mu(t,\vec{x})$ will be splitted into $V^0(t,\vec{x})$ and $V^i(t,\vec{x})$, where each of these are to be taken now tensors living on a spatial slice at the instant $t$: first one a scalar and the second one a vector field on the spatial slice. Similarly a rank-2 space-time tensor will be decomposed into a scalar, vector and tensor on a spatial slices. We refer the reader to standard texts on space-time splitting for further details, for example \citep{MFB},\citep{baumann}.\\

After the space-time decomposition, where everything is described on a spatial slice which we will take to be a $\pr$-manifold; we will perform our new decompositions described above. In the first subsection we first describe how the metric and energy-momentum, and various expressions involving them decompose under our theorems. After this we write down cosmological perturbation equations with the new decomposition. Finally we end the section with a discussion of the decomposition of gauge transformations and the forming of new gauge invariant quantities.

\subsection{Decomposition of perturbative fields}
 First we describe how perturbations and various expressions that appear in the equations of motions that involve them decompose.  As usual we will perform a space-time splitting: In the cosmological coordinates we will work in, this will simply mean writing out the time and space indices explicitly. Thus we write down the FRLW background metric with perturbations as
\begin{equation} \label{eq:full metric}
(\bar{g}_{\mu\nu}+h_{\mu \nu})dx^\mu dx^\nu = - (1+2 \Phi) dt^2 + 2 a(t) N_i dx^i dt + a^2(t) \lp (1-2\Psi) \delta_{ij} + 2 \gamma_{ij} \rp dx^i dx^j \ ,
\end{equation}
where the perturbations $\Phi,N_i,\Psi,\gamma_{ij}$ have space-time dependency on the coordinates $(t,x^i)$. We will take the energy-momentum tensor to be such that
\begin{equation}
T^0_0=-\lp \bar{\rho}(t) +  \rho \rp \cg T^0_i= \lp \bar{\rho}(t) + \bar{p} \rp v_i \cg
T^i_j = \delta_{ij} \lp \bar{p}(t) +  p \rp + \Sigma_{ij} 
\end{equation}
where $\bar{\rho}(t)$ and $\bar{p}(t)$ are background quantities and the perturbations $\rho,p,v,\Sigma$ has again space-time dependency. After this space-time splitting, perturbations of the Einstein tensor $\delta G_{00}(t,x^i), \delta G_{0i}(t,x^i)$ and $\delta G_{ij}(t,x^i)$ now can be considered as a scalar, one-form and contravariant rank-2 symmetric tensor respectively on each $t=\mbox{constant}$ slice. Note that each of these slices are flat, and we let them be $\pr$-manifolds with boundary. Then our next job is to decompose all the types of tensors on any slice. For the scalar part, in place of the Hodge-Morrey theorem (\ref{thm:HM}) we will use the decomposition 
\begin{equation}\label{eq:sc_dec}
f=f^t+f^h \gwg \bt f^t=0 \gag \dd d f^h=0 \ .
\end{equation}
This decomposition is unique, since by Theorem 3.4.10 in \citep{Gunter} there exists a unique solution to the boundary value problem
\begin{equation}
\dd d f^h=0 \gwg \bt f^h= \bt f 
\end{equation}
for $f^h$.
For the one-form we will use the Hodge-Morrey theorem and the fact that our 3-manifold has trivial cohomology. Then one can write the decomposition as
\begin{equation}\label{eq:vec_dec2}
v = d(v^t+v^h) + v^n 
\end{equation}
where
\begin{equation}
\bt v^t=0 \cg \dd d v^h=0 \cg \dd v^n=0 \gwg \bn v^n=0 \ .
\end{equation}
Using these, we will decompose our perturbations as follows: Each of the scalars such as $\Phi$ or $\rho$ will be decomposed as \eqref{eq:sc_dec} whenever needed, simply by adding the superscripts. We decompose one-forms as
\begin{equation}
N_i= \pr_i \psi + N_i^n \gag v_i=\pr_i v + v_i^n
\end{equation}
where scalars will be again decomposed as \eqref{eq:sc_dec} whenever needed and
\begin{equation}
\dd N^n=0 \cg \bn N^n=0 \cg \dd v^n=0 \gag \bn v^n=0 \ .
\end{equation}
Tensors will be decomposed as 
\begin{align}
\gamma_{ij}&= \dmd_{ij} \hgamma + \pr_{(i} \gamma^{nd}_{j)} + \nabla_i \nabla_j \gamma^h + \gamma^{TTA}_{ij} \label{eq:gamma-dec} ,\\
\Sigma_{ij}&= \dmd_{ij} \hSigma + \pr_{(i} \Sigma^{nd}_{j)} +\nabla_i \nabla_j \Sigma^h + \Sigma^{TTA}_{ij} \ ,
\end{align}
and each part satisfies the conditions given in theorem \ref{thm:SVT} and theorem \ref{thm:sec_dec}. Here we note the difference between e.g. $\gamma^{nd}$ and $N^n$: $\gamma^{nd}$ needs to satisfy $\bn d \gamma^{nd}=0$ while no such condition exists for $N^n$. 

Scalars, one-forms and tensors introduced above will enter into the linearized Einstein equations in various ways. Most non-trivial of these are the expressions of the form $\pr_{(i}N_{j)}$. Using the one-form decomposition \eqref{eq:vec_dec2} one can write for the traceless part of this
\begin{equation}
\lp \pr_{(i} N_{j)} \rp^{\mbox{t'less}}= \dmd_{ij} (\psi^t+\psi^h) + \pr_{(i} N_{j)}^n \ .
\end{equation}
Here first two terms belong to the t-part and h-part of the tensor decomposition respectively. Then one is left with the decomposition of the $\pr_{(i} N_{j)}^n$ part. Note that this is not of the $nd$-part type of the rank-2 tensor decomposition since it does not satisfy $\bn d N^{nd}=0$ in general. Now we argue the following:
\begin{thm}\label{thm:pr_N}
Any one form $N^{n}$ on a $\pr$-manifold with trivial cohomology such that $\dd N^n=0$ and $\bn N^n=0$ can be decomposed as
\begin{equation}
N^n=N^{nd} + \bar{N}^n
\end{equation}
such that 
\begin{align}
& \dd N^{nd}=0 \cg \bn N^{nd} =0 \cg \bn d N^{nd}=0 \ , \\
& \dd \bar{N}^n=0 \cg \bn \bar{N}^n =0 \cg d\dd d \bar{N}^n=0 \ . \label{eq:Nbar}
\end{align}
\end{thm}
This theorem can be proven by three successive application of the corollary \ref{cor:dw}. Now if we write
\begin{equation}\label{eq:pr_i-NT_j}
\pr_{(i} N_{j)}^n= \dmd_{ij} N^{th} + \pr_{(i} N^{nd}_{j)} +  N_{ij}^{TT} \ , 
\end{equation}
the fact that $d\dd d \bar{N}^n=0$ implies $\pr^2\pr^2N^{th}=0$, in addition to the conditions in theorem \ref{thm:SVT}, namely
\begin{equation}
\bt N^{th}=0 \ , g^{ij}N_{ij}^{TT}=0 \ , \nabla^i N_{ij}^{TT}=0 \ .
\end{equation}
Note that the full decomposition of $\pr_{(i} N_{j)}$ becomes
\begin{equation}
\pr_{(i} N_{j)}= \frac{g_{ij}}{m} \pr^2 \psi^t + \dmd_{ij} (\psi^t+N^{th}) + \pr_{(i} N^{nd}_{j)} + \nabla_i \nabla_j (\psi^h+N^h) + N_{ij}^{TTA} \ ,
\end{equation}
where we also have performed the second decomposition.

\subsection{Decomposition of linearized Einstein equations}
Now we use the decompositions described above in the Einstein equations $G_{\mu \nu}=T_{\mu \nu}$ for the perturbed metric and the energy-momentum tensor. We use the coordinates $(t,x^i)$ described in \eqref{eq:full metric}. $(t,t)$ component of the equation becomes
\begin{equation}
-6H\dot{\Psi}+ 2\frac{\pr^2 \Psi}{a^2}-2H \frac{\pr^2\psi}{a}+ \frac{2}{3a^2} \pr^2 \pr^2 \hgamma = \rho + 6H^2 \Phi \ .
\end{equation}
Note that this scalar equation can be further decomposed into harmonic and t-parts. For the $(t,i)$ components expressions 
\begin{equation}
\pr^2 N^n_i \gag \pr^2 \gamma^{nd}_i
\end{equation}
will be needed. Using \eqref{eq:pr_i-NT_j} we note
\begin{equation}\label{eq:pr2NTj}
\pr^2 N_j^n= \pr^2 N^{nd}_j + \frac{4}{3} \pr_j \pr^2 N^{th} \ .
\end{equation}
Note that this has the correct form for the one-form decomposition \eqref{eq:vec_dec2}: $\pr^2 N^{nd}_j \sim (\dd d N^{nd})_j$ and
\begin{equation}
\dd (\dd d N^{nd})=0 \gag \bn d N^{nd}=0 \rightarrow \bn (\dd d N^{nd}) = 0 \ ,
\end{equation}
thus $\pr^2 N^{nd}_j$ is of the $v^n_i$ type in the one-form decomposition. The last implication arrow can be proved by using identities \eqref{eq:star-t identity} and \eqref{eq:star-d identity}. The second part in \eqref{eq:pr2NTj} is then of $\pr_i v^h$ type. With a similar argument we see $\pr^2 \gamma^{nd}_i$ is only of the $v^n_i$ type. Using these then the $\pr_i (v^t+v^h)$ part of the $(t,i)$ equation becomes
\begin{equation}
\pr_i(\dot{\Psi}+ H\Phi+ \frac{1}{3} \pr^2\dot{\hgamma}-\frac{1}{3a}\pr^2 N^{th}-\dot{H}v)=0 \ ,
\end{equation} 
whereas $v^n$ part becomes
\begin{equation}
-\frac{1}{2a} \pr^2 N^{nd}_j + \frac{1}{2} \pr^2 \dot{\gamma}^{nd}_j-2\dot{H}v_i^n=0 \ .
\end{equation}
For the $(i,j)$ part, apart from what we have already covered, we only need the decomposition of the following expression:
\begin{equation}
-\pr^2 \gamma_{ij} + 2 \pr_k \pr_{(i} \gamma_{j)k} - \frac{2}{3} \delta_{ij} \pr_m \pr_k \gamma_{mk} = \frac{1}{3} \dmd_{ij} \pr^2 \hgamma - \pr^2 \gamma^{TTA}_{ij} \ ,
\end{equation}
where we have simply plugged in \eqref{eq:gamma-dec}. Note that $\pr^2 \gamma^{TTA}_{ij}$ is TTA-type and the equality above provides the proper decomposition. Using this we write five parts of the $(i,j)$ equation: one trace equation and three coming from the decomposition of the traceless part of the equation. Trace part is
\begin{equation}
(\pr_t+H) \left[ \pr^2(a\psi)+ 3 a^2(\dot{\Psi}+H\Phi)\right] + \pr^2(\Phi-\Psi)-\frac{1}{3} \pr^2 \pr^2 \hgamma = \frac{3a^2}{2} (p-2\dot{H}\Phi) \ .
\end{equation}
$\dmd_{ij}$ part is
\begin{equation}
\dmd_{ij} \left[ \frac{1}{a} (\pr_t + 2H) \lp a\dot{\hgamma} - \psi^t-N^{th} \rp + \frac{1}{a^2} \lp \Psi^t + \frac{(\pr^2\hgamma)^t}{3} \rp- \frac{\Phi^t}{a^2}-\hSigma \right]= 0 \ .
\end{equation}
nd-part of the $(i,j)$ equation is 
\begin{equation}
(\pr_t+H) \left[ a^2 \pr_{(i} \dot{\gamma}^{nd}_{j)} - a \pr_{(i} N^{nd}_{j)} \right] = a^2 \pr_{(i} \Sigma^{nd}_{j)} \ .
\end{equation}
Lastly we find the h-part and TTA-part of the $(i,j)$ equation. We find
\begin{equation}
(\pr_t+H) \lp a^2 \gamma^{TTA}_{ij}-a N^{TTA}_{ij} \rp -a^2 \Sigma^{TTA}_{ij}= 0 
\end{equation}
and
\begin{equation}
\nabla_i \nabla_j \lp (\pr_t+H) \lp a^2 \dot{\gamma}^{h}-a \mathcal{N}^{h} \rp -a^2 \Sigma^{h}+   \frac{(\pr^2\gamma)^h}{3} -(\pr_t+H)(a\psi^h)+(\Psi-\Phi)^h \rp = 0 \ .
\end{equation}
\paragraph{Summary: Equations of motion}
\subparagraph{Scalars}
\begin{flalign}
& 6H^2 \Phi+6H\dot{\Psi}- 2\frac{\pr^2 \Psi}{a^2}- \frac{2}{3a^2} \pr^2 \pr^2 \hgamma+2H \frac{\pr^2\psi}{a} + \rho =0 \label{eq:scalar-1} \ ,&\\
&\pr_i( H\Phi+ \dot{\Psi}+ \frac{1}{3} \pr^2\dot{\hgamma}-\frac{1}{3a}\pr^2 N^{th}-\dot{H}v)=0 \ , \label{eq:scalar-2} &\\
&(\pr_t+H) \left[ 3 a^2(\dot{\Psi}+H\Phi) + \pr^2(a\psi)\right] + \pr^2(\Phi-\Psi) -\frac{1}{3}\pr^2 \pr^2\hgamma= \frac{3a^2}{2} (p-2\dot{H}\Phi) \ ,& \label{eq:scalar-3} \\
&\dmd_{ij} \left[ - \Phi^t +  \Psi^t + \frac{(\pr^2\hgamma)^t}{3} +  (\pr_t + H) \lp a^2\dot{\hgamma} - a\psi^t-a N^{th} \rp -a^2\hSigma \right]= 0 \ ,& \label{eq:scalar-4} \\
&\nabla_i \nabla_j \lp (\Psi-\Phi)^h + \frac{(\pr^2\gamma)^h}{3} + (\pr_t+H) \lp a^2 \dot{\gamma}^{h}-a N^{h} -a\psi^h\rp -a^2 \Sigma^{h} \rp = 0 \ .& \label{eq:scalar-5}
\end{flalign}
\subparagraph{Vectors}
\begin{flalign}
&\frac{1}{2} \pr^2 \dot{\gamma}^{nd}_j-\frac{1}{2a} \pr^2 N^{nd}_j -2\dot{H}v_i^n=0 \ , & \\
&(\pr_t+H) \left[ a^2 \pr_{(i} \dot{\gamma}^{nd}_{j)} - a \pr_{(i} N^{nd}_{j)} \right] -a^2 \pr_{(i} \Sigma^{nd}_{j)}=0 & .
\end{flalign}
\subparagraph{Tensors}
\begin{flalign}\label{eq:TTA-eom}
&(\pr_t+H) \lp a^2 \gamma^{TTA}_{ij}-a N^{TTA}_{ij} \rp -a^2 \Sigma^{TTA}_{ij}= 0 \ . &
\end{flalign}
Let us now take a moment to reflect on the degree of freedom counting: we start with the metric. It has 10 scalars, 4 of which is t-type i.e. $\bt f$=0 for them; 5 of them is h-type, they are solutions to the Laplace equation and the last one is th-type it satisfies $\bt f=0$ and $\pr^2 \pr^2 f=0$. Note that only a combination of $t$ and $h$ type gives one a complete ``arbitrary" function, recovering the usual 4-scalars of the metric. The scalars $N^{th}$ and $N^h$ needs to be solving some equations so they should not be included in a degree of freedom counting, where freedom means a full arbitrary function. These can be considered to be ``stealing" from the vector degrees of freedom. 

In the tensorial equation \eqref{eq:TTA-eom}, the appearance of $N^{TTA}_{ij}$ looks preplexing. Note that this not a most general $TTA$ type tensor, it comes from the expression $\pr_{(i} N^n_{j)}$ and thus it needs to satisfy extra conditions such as those implied by \eqref{eq:Nbar}. We leave the full extraction of degrees of freedom to a future work, but note that this extraction would decompose \eqref{eq:TTA-eom} further into
\begin{flalign}
&(\pr_t+H) \lp a^2 \pr_{(i} \gamma^{TTAV}_{j)}-a \pr_{(i} \bar{N}^n_{j)} \rp -a^2 \pr_{(i}\Sigma^{TTAV}_{j)}= 0 \ , \label{eq:TTAV1}\\
&(\pr_t+H) \lp a^2 \gamma^{TTA-rest}_{ij} \rp -a^2 \Sigma^{TTA-rest}_{ij}= 0 \ , \label{eq:TTAV2}
\end{flalign}
where $\bar{N}^n_j$ was introduced in theorem \ref{thm:pr_N}, and $TTAV$ parts are the parts of respective tensors in the form of $\pr_{(i} \bar{N}^n_{j)}$. Second equation then reflects the remaining parts that cannot be expressed in terms of derivatives of a vector.

\subsection{Gauge transformations}
It is of interest to see how each component of the metric and the energy momentum tensor varies under an infinitesimal coordinate transformation. The main reason is that one would like to know if a perturbation is a mere coordinate transformation artifact and if so simplify the equations by eliminating these artifacts. Writing down
\begin{equation}
\delta_\xi h_{\mu\nu} = \lie_\xi \bar{g}_{\mu \nu}
\end{equation}
and decomposing $\xi=(\xi^0,\xi^i)$ ,
\begin{equation}
\xi_i=\pr_i(\xi^t+\xi^h) + \xi^n_i
\end{equation}
and
\begin{equation}
\pr_{(i}\xi_{j)}^T= \dmd_{ij} \phi^{th} + \pr_{(i} \xi^{nd}_{j)}+ \pr_i\pr_j \phi^h + \xi^{TTA}_{ij}
\end{equation}
using theorem \ref{thm:sec_dec} and equation \eqref{eq:pr_i-NT_j}, we see that the scalars transform as\\
\begin{minipage}[b]{0.5\linewidth}
\begin{align}
& \delta \Phi = -\dot{\xi_0} \ , \\
& \delta \Psi = H\xi_0- \frac{\pr^2 \xi^t}{3a^2} \ , \\
& a \pr_i \delta \psi = \pr_i (\xi_0+ \dot{\xi}-2H\xi) \ , \\
& a^2 \dmd_{ij} \delta\hgamma = \dmd_{ij} (\xi^t+\phi^{th}) \ , \\
& a \dmd_{ij} \delta N^{th} = (\pr_t -2H) \dmd_{ij} \phi^{th} \ , 
\end{align}
\end{minipage}
\begin{minipage}[b]{0.5\linewidth}
\begin{align}
& a^2 \pr_i \pr_j \delta \gamma^{h} = \pr_i \pr_j (\xi^h + \phi^{h}) \ , \\
& a \pr_i \pr_j N^{h}=(\pr_t-2H)\pr_i \pr_j \phi^{h} \ , \\
& \delta \rho=-\dot{\bar{\rho}} \xi_0 \cg \delta p = - \dot{\bar{p}} \xi_0  \\
& \pr_i \delta v= \pr_i \xi_0  \ , \\
& \delta \hSigma=0 \cg \delta \Sigma^{h}=0
\end{align}
\end{minipage}\\
whereas vectors and tensors transform as \\
\begin{minipage}[b]{0.5\linewidth}
\begin{align}
& a \pr_{(i}\delta N^{nd}_{j)}=(\pr_t-2H) \pr_{(i} \xi^{nd}_{j)} \ , \\
& a^2 \pr_{(i}\delta \gamma^{nd}_{j)}=\pr_{(i} \xi^{nd}_{j)} \ , \\
& \delta v_i^n=0 \ , 
\end{align}
\end{minipage}
\begin{minipage}[b]{0.5\linewidth}
\begin{align}
& a^2 \delta \gamma^{TTA}_{ij}= \xi^{TTA}_{ij} \ , \label{eq:gamma-TTA-gauge-tfm}\\
& a \delta N^{TTA}_{ij}= (\pr_t-2H)\xi^{TTA}_{ij} \ , \\
& \delta \Sigma^{TTA}_{ij}=0 \ .
\end{align}
\end{minipage}\\
Note that here it looks like $\gamma^{TTA}_{ij}$ is gauge transforming, however upon a further decomposition of it into a $\pr_{(i}V_{j)}$ type expression and the rest, some gauge invariant part should emerge as was discussed at the end of the previous subsection. 

Using the information above gauge invariant combinations can be formed, but as removing the derivatives will introduce ambiguities unlike the standard case, these terms will typically involve more derivatives. For example analogues of Bardeen fields \citep{Bardeen}, which are gauge invariant, will be
\begin{align}
& \pr^2 \pr^2 \dot{\Psi}^t_B= \pr^2 \pr^2 \lp \dot{\Psi}^t + \frac{(\pr^2 \hgamma)^t}{3} + \pr_t (H a^2 \hgamma)- \pr_t(Ha \psi^t) \rp \ ,\\
& \pr_i \pr_j \Phi^h_B = \pr_i \pr_j \lp \Phi^h-\pr_t(a\psi^h)+ \pr_t (a^2 \dot{\gamma}^{h}-\pr_t (a^2 \pr_t(a^{-1} N^{h}))) \rp \ . \label{eq:PhiB}
\end{align}

\section{Application to some cosmological scenarios}\label{sec:app-to-som}
Having set the rules for the new cosmological perturbation theory, we would like to see how some specific scenarios work with it. We will focus on scalar equations of motion, leaving the study of vectors and tensors to future work. In the first subsection we study the single field inflation, and reproduce two copies of the Mukhanov-Sasaki equation: exactly in the same form for a t-type scalar and a modified version for h-type scalar. In the second subsection we reproduce a Weinberg adiabatic mode like solutions, namely by letting all fields except Lagrange multipliers to be pure gauge we solve the cosmological perturbation equations, and obtain a non-trivial gauge invariant solution for an arbitrary scale factor $a(t)$.

\subsection{Single field inflation}
We will first consider the case where a single field with a space independent background constitutes the matter, where action takes the form
\begin{equation}
S= \int d^4x \sqrt{-g} \lp R- \frac{1}{2} g^{\mu \nu} \pr_\mu \tilde{\varphi} \pr_\nu \tilde{\varphi}-V(\tilde{\varphi}) \rp \ .
\end{equation} 
For this action components of the energy momentum tensor as it appears in the Einstein equation $G_{\mu\nu}=T_{\mu\nu}$ has the form
\begin{align}
\bar{\rho}&= \frac{1}{2} \lp \frac{\dot{\bvphi}^2}{2}+V(\bvphi) \rp \cg  
\rho = \frac{\dot{\bvphi}}{2} \dot{\varphi} + \frac{V'(\bvphi)}{2} \varphi + 2 \dot{H} \Phi \ , \\
\bar{p}&= \frac{1}{2} \lp \frac{\dot{\bvphi}^2}{2}-V(\bvphi) \rp \cg  
p = \frac{\dot{\bvphi}}{2} \dot{\varphi} - \frac{V'(\bvphi)}{2} \varphi + 2 \dot{H} \Phi \ ,\\
v &= - \frac{\varphi}{\dot{\bvphi}} \cg v_i^T=0 \cg \Sigma_{ij}=0 \ .
\end{align}
Following \citep{weinbergcosmology} let us choose a gauge where $\varphi=0$, $\pr_i \pr_j N^{h}=0, \pr_i \pr_j \gamma^{h}=0, \dmd_{ij}N^{th}=0,\dmd_{ij} \hgamma=0$. \footnote{Note that this can be chosen, but the gauge is not completely fixed.} Then the first two scalar equations \eqref{eq:scalar-1} and \eqref{eq:scalar-2} become 
\begin{align}
& (3H^2+\dot{H})\Phi+ 3H\dot{\Psi}- \frac{\pr^2\Psi}{a^2} + H \frac{\pr^2 \psi}{a} = 0 \ , \\
& \pr_i \lp H\Phi + \dot{\Psi} \rp =0 \label{eq:phi-psi} \ .
\end{align}
In addition to these, following \citep{weinbergcosmology} we will use the time component of the energy-momentum conservation equation which can be written as
\begin{equation}
\dot{H} \dot{\Phi} + (\ddot{H} + 6 H \dot{H}) \Phi + \dot{H} \lp 3 \dot{\Psi} + \frac{\pr^2\psi}{a} \rp = 0 \ .
\end{equation}
Note that \eqref{eq:phi-psi} can be written as
\begin{equation}
H\Phi^t + \dot{\Psi}^t=0 \gag H\Phi^h + \dot{\Psi}^h=A(t) 
\end{equation}
where $A(t)$ is an arbitrary space-independent field. Using this and other two equations above one can write
\begin{align}
& \ddot{\Psi}^t + \lp \frac{\ddot{H}}{\dot{H}}- 2 \frac{\dot{H}}{H}+ 3H \rp \dot{\Psi}^t - \frac{(\pr^2\Psi)^t}{a^2}=0 \ , \\
& (\ddot{\Psi}^h-\dot{A}) + \lp \frac{\ddot{H}}{\dot{H}}- 2 \frac{\dot{H}}{H}+ 3H \rp (\dot{\Psi}^h -A) - \frac{(\pr^2\Psi)^h}{a^2}=0 \ . \label{eq:MS1}
\end{align}
These are two copies of the Mukhanov-Sasaki equation.\footnote{See (10.1.35) in \citep{weinbergcosmology}, originally produced in \citep{mukhanov},\citep{sasaki}.} Note that $\Psi^t$ piece works exactly the same as the standard case with vanishing boundary, whereas the harmonic part $\Psi^h$ couples with the arbitrary field $A(t)$. This arbitrary part is expected to be fixed by the boundary conditions.

\subsection{An alternative adiabatic-like mode}
Next we study a case where the role of boundary conditions is more subtle. We will find some solutions with a procedure motivated by Weinberg's adiabatic mode solutions \citep{weinbergadiabatic} around an arbitrary FRLW background. Note that in Weinberg's argument, one starts from spatial independent, i.e. zero momentum, solutions and takes the limit of non-zero momentum while keeping the assumption of vanishing of fields at spatial infinity. Strictly speaking there exists no non-trivial solution that is space independent and vanishing at infinity, and removing this condition at infinity makes the standard scalar-vector-tensor decomposition unavailable. 

In the following, using the decomposition described above we will argue the existence of some gauge-invariant solutions to the perturbed Einstein equations whatever the constituents of the universe. These solutions then have the potentiality to connect the inflationary era to the post-last scattering era, as was the motivation of Weinberg's paper.

We propose the following scheme: In the perturbed Einstein equations we will let all the fields (including matter) except the Lagrange multipliers to be pure gauge and then we solve the equations for these. We will only consider the scalar equations, which have decoupled from vector and tensor equations at this stage. Note that scalar Lagrange multipliers are $\Phi^{t,h},\pr_i \psi^{t,h}, N^{th}, N^{h}$ . One of the scalar equations of motion becomes
\begin{equation}
\pr_i \lp H (\dot{\xi}_0^t+\Phi^t) \rp =0
\end{equation}
implying $\Phi^t=-\dot{\xi}^t_0$ since $\bt \Phi^t=0$. Using this inside the other equations then enforces $\Phi^h,\psi^t$ to be pure gauge. Remaining equations then reads
\begin{align}
& \pr_i \pr^2 \lp a N^{th} -(\pr_t-2H) \phi^{th} \rp = 0 \ , \\
& \dmd_{ij} (\pr_t+H) \lp a N^{th} -(\pr_t-2H) \phi^{th} \rp = 0 \ , \\
& \pr_i \pr_j (\pr_t+H) (-a\psi^h+(\pr_t-2H)\xi^h-aN^{h} + (\pr_t-2H)\phi^{h})= 0 \ .
\end{align}
Note that the equation $\dmd_{ij} \lambda=0$ is solved by
\begin{equation}
\lambda=A(t)r^2 + B_i(t)x^i + C(t) \ .
\end{equation}
Using this, one can then solve the remaining equations of motion as
\begin{align}
& a N^{th} = a \delta_g N^{th} + A(t)(r^2-R^2) \ , \\
& \pr_i \pr_j \pr_t (a \psi^h + a N^{h}) = \pr_i \pr_j \pr_t \delta_g(a \psi^h + a N^{h}) - \frac{\pr_i \pr_j \alpha^h(x)}{a^2} \ .
\end{align}
where $\delta_g$ means pure gauge, $A(t)$ is arbitrary, $\alpha^h(x)$ is arbitrary harmonic and function of only the spatial coordinates. Using this in the formula \eqref{eq:PhiB} for the gauge invariant quantity $\pr_i \pr_j \Phi^h_B$ we see
\begin{equation}\label{eq:Gauge-inv-ad}
\pr_i \pr_j \Phi^h_B = \frac{\pr_i \pr_j \alpha^h(x)}{a^2(t)} \ .
\end{equation}
Note that this solution exists irrespective of what the scale factor $a(t)$ is. In an expanding universe these are diluted harmonics, which can be fixed by spatial boundary conditions. Physical interpretation of them should depend on these boundary conditions, but from the discussion above it is clear that they are not gauge artifacts.

\section{Discussion}
In the work above we have developed a scalar-vector-tensor decomposition of symmetric rank-2 tensors on Ricci flat manifolds with boundary and used this decomposition to rewrite the cosmological perturbation theory in a way that allows for general boundary conditions. In deriving the decomposition theorem \ref{thm:SVT} we have utilized the Hodge-Morrey decomposition of forms but at later stages we needed to assume trivial cohomology and Ricci flatness for the $\pr$-manifold we are working on. These assumptions are good enough for a study of fields on ball-like spaces, but it would be interesting to generalize this to more general conditions. For example one might want to solve the Einstein equations inside and outside of a ball and match the solutions on the surface. Assuming a vanishing boundary condition outside of the ball, one would have the topology of $S^n$ with a hole. In this case the assumption of trivial cohomology no longer holds and one would need to rerun the arguments leading to the decomposition of theorem \ref{thm:SVT}. The matching of fields on the boundary would fix the arbitrary coefficients/fields such as the coefficients of spherical harmonics in \eqref{eq:Gauge-inv-ad} or the arbitrary field $A(t)$ in \eqref{eq:MS1}, and allow us to have a better physical understanding of these solutions. Note that the important advantage of our procedure is that it is in principle built on a theorem that holds for a very general case and it is clear which steps should be restudied in order to generalize it. 

We also would like to note that in its current form our decomposition quite possibly is not the ideal decomposition one might want to have in order to study more general boundary conditions, there exists still some mixing of different types of terms in the perturbation or gauge transformation equations, for example mixing of $N^{th}$ with t-type scalars in \eqref{eq:scalar-4}, or mixing of $\pr_{(i} \bar{N}^n_{j)}$ with TTA-type tensors as was explained above equation \eqref{eq:TTAV1}. Further decompositions that allow for these are possible of course but the number of types of fields increase with each such decomposition. It would be most desirable to put this clutter of decompositions into a more systematic procedure. One can appeal to a more geometric investigation, but the role of boundary conditions may not be very trivial in this type of discussion. Another possibility that might be more practical is to use a polynomial expansion in the spirit of \citep{Chodos} and make use of tensor spherical harmonics for the whole of discussion.

Finally we would like to note that we expect this formalism to be useful not only for problems where a matching between a bounded region and outside of it is made. For example in Weinberg's argument for the existence of adiabatic modes \citep{weinbergadiabatic} one starts from the spatially homogeneous-i.e. zero momentum solutions and connects this to the non-zero momentum case. This should also call for connecting the decomposition with non-vanishing boundary conditions to the decomposition with vanishing boundary conditions. In this sense it would be interesting to investigate if this decomposition is useful for the discussions of asymptotic symmetries, where sometimes some degrees of freedom are allowed to exist at the asymptotics, such as supertranslations and superrotations for asymptotically flat spacetimes.

\acknowledgments
I would like to thank Emre Onur Kahya of İTÜ, Olaf Lechtenfeld of Leibniz University and Bayram Tekin of ODTÜ for hosting me throughout the project and discussions. I would like to thank Bayram Tekin also for pointing out the paper of Straumann \citep{Straumann}. Author has been supported by: TÜBİTAK Grant No.122R070, TÜBİTAK-2219 Fellowship and ODTÜ BAP DOSAP-10763 throughout the project.

\appendix
\section{Some Mathematical Facts}
\subsection{Some exterior algebra expressions}
On any n-dimensional smooth manifold, for any $\omega$ k-form, $\eta$ m-form, $x^i$ where $i=1,\cdots,n$ local coordinates
\begin{flalign}
& (d\omega)_{i_1 \cdots i_{k+1}}= (k+1) \pr_{\lb i_1 \right.} \omega_{ \left. i_2 \cdots i_{k+1} \rb} \ , \\
&dx^{i_1} \wedge \cdots \wedge dx^{i_k} = k! \ dx^{[i_1} \otimes \cdots dx^{i_k]} \sgc \\
& \omega \wedge \eta = \frac{(k+m)!}{k! m!} \omega_{[\ind{i}{1}{k}} \eta_{\ind{i}{k+1}{k+m}]}  dx^{i_1} \otimes \cdots dx^{i_n}  \sgd 
\end{flalign}
On any Riemannian manifold with Riemannian volume form $\epsilon$
\begin{flalign}
& \dd \omega = (-1)^{nk+n+1} *d* \omega \ , \\
& \lp *w \rp _\ind{i}{k+1}{n} = \frac{1}{k!} {\epsilon^\ind{j}{1}{k}} _\ind{i}{k+1}{n} w_\ind{j}{1}{k} \sgc \\
&\epsilon^{\ind{i}{1}{p}\ind{k}{p+1}{n}} \epsilon_{\ind{i}{1}{p}\ind{j}{p+1}{n}}= (n-p)! p! \, \delta^{[k_{p+1}}_{j_{p+1}} \cdots \delta^{k_{n}]}_{j_{n}} \sgd
\end{flalign}

\subsection{Some facts on de Rham Cohomology groups} \label{sec:deRham Groups}

\textbf{Poincare Lemma}: If $M$ is contractible to a point then $H^k_{dR}(M)=0$ for $k=1, \cdots,n$. Note that $R^n$ is contractible to a point.\\
\hfill\\
\textbf{Some Boundary Value Problem Theorems} \\
Here we restate Corollary 3.2.6 in \citep{Gunter} on $\pr$-manifolds, simplified for trivial cohomology and under some special circumstances:
\begin{cor}\label{cor:dw}
 Let $w$ be a k-form where $k \ge 1$. Then the boundary value problem 
\begin{equation}
dw=\chi \cg \dd w= 0 \cg \bn w=0
\end{equation}
has a unique solution if $d\chi=0$.
\end{cor}
\begin{cor}\label{cor:ddw}
Let $w$ be a k-form where $k \ge 1$. Then the boundary value problem 
\begin{equation}
dw=0 \cg \dd w= \rho \cg \bn w=0
\end{equation}
has a unique solution if 
\begin{itemize}
\item $\dd \rho=0$ and $\bn \rho=0$ for $k>1$ and
\item $ \ket \rho, \kappa \bra=0$ for all $\kappa \in \R$ for $k=0$.
\end{itemize}

\end{cor}

\bibliographystyle{plain}
\bibliography{bibliography}

\end{document}

%% file: globaldef.tex
\newcommand{\rp}{\right)}
\newcommand{\lp}{\left(}
\newcommand{\rb}{\right]}
\newcommand{\lb}{\left[}
\newcommand{\rcb}{\right\rbrace}
\newcommand{\lcb}{\left\lbrace}

\newcommand{\bra}{\right>}
\newcommand{\ket}{\left<}

\newcommand{\ra}{\rightarrow}

\newcommand{\pr}{\partial}

\newcommand{\lie}{\mathcal{L}}

\DeclareMathOperator{\R}{\mathbb{R}}

\newcommand{\cg}{\, , \quad}
\newcommand{\sgd}{\, .}
\newcommand{\sgc}{\, ,}
\newcommand{\gag}{\quad \mbox{and} \quad}
\newcommand{\gwg}{\quad \mbox{where} \quad}

%% file: commenting.tex
\newcommand{\listoflan}{List of Possible Language Mistakes}
\newlistof[chapter]{lan}{lan}{\listoflan}

\newcommand{\listtoexplore}{List of Things to Explore Further}
\newlistof[chapter]{expl}{expl}{\listtoexplore}

\newcommand{\listofquestions}{List of Questions}
\newlistof[chapter]{que}{que}{\listofquestions}

\newcommand{\listoflattech}{Latex Technical Problems}
\newlistof[chapter]{lattech}{lattech}{\listoflattech}

